\documentclass{aa}

\newbox\grsign \setbox\grsign=\hbox{$>$} \newdimen\grdimen \grdimen=\ht\grsign
\newbox\simlessbox \newbox\simgreatbox \newbox\simpropbox \newbox\wtildebox 
\setbox\simgreatbox=\hbox{\raise.5ex\hbox{$>$}\llap
     {\lower.5ex\hbox{$\sim$}}}\ht1=\grdimen\dp1=0pt
\setbox\simlessbox=\hbox{\raise.5ex\hbox{$<$}\llap
     {\lower.5ex\hbox{$\sim$}}}\ht2=\grdimen\dp2=0pt
\def\simgreat{\mathrel{\copy\simgreatbox}}
\def\simless{\mathrel{\copy\simlessbox}}

\newcommand{\as}{ A$\&$A }

\newcommand{\Lsun}{\mbox{$L_\odot$}}
\newcommand{\Msun}{\mbox{$M_\odot$}}

\newcommand{\Rsun}{\mbox{$R_\odot$}}
\newcommand{\vinf}{\mbox{$v_\infty$}}
\newcommand{\vesc}{\mbox{$v_{\rm esc}$}}
\newcommand{\mdot}{\mbox{$\dot{M}$}}
\newcommand{\Mdot}{\mbox{$\dot{M}$}}

\newcommand{\Teff}{\mbox{$T_{\rm eff}$}}

\newcommand{\Rs}{\mbox{$R_{\rm s}$}}
\newcommand{\Ts}{\mbox{$T_{\rm s}$}}
\newcommand{\Msunyr}{\mbox{$\Msun~{\rm yr}^{-1}$}}
\newcommand{\kms}{\mbox{km~s$^{-1}$}}

\begin{document}

\title{An explanation for the curious mass loss history of massive stars:\\
       from
       OB stars, through Luminous Blue Variables to Wolf-Rayet stars }

\author{H. J. G. L. M. Lamers  \inst{1,2}
  and T. Nugis \inst{3}}

\offprints{H.J.G.L.M. Lamers \tt{lamers@astro.uu.nl}}

\institute{{Astronomical Institute, 
              University of Utrecht, 
              Princetonplein 5, NL-3584CC, Utrecht,
              the Netherlands}
          \and  {SRON Laboratory for Space Research,
              Utrecht, the Netherlands}
          \and {Tartu Observatory, 61602 T\~oravere, Estonia}
          }    

\date{Received date ; accepted date}

\authorrunning{H.J.G.L.M. Lamers \& T. Nugis}
\titlerunning{Explaining the mass loss history of massive stars} 

\abstract{The stellar winds of massive stars show large changes in 
mass-loss rates and terminal velocities during their evolution from
O-star through the Luminous Blue Variable 
phase to the Wolf-Rayet phase. 
The luminosity remains approximately unchanged during these phases.
These large changes in wind properties are explained
in the context of the radiation driven wind theory, of which we
consider four different models. They are due to 
the evolutionary changes in radius, gravity and surface composition 
and to the change from optically thin (in continuum) line driven
winds to optically thick radiation driven winds.
}

\maketitle
%
%
\section{Introduction}

%
%
 In this paper we present an explanation for the drastic changes
  in mass loss rate and wind velocity during the evolution of massive
  stars.

The stellar winds of massive stars show a large variation in 
mass-loss rates and terminal velocities during their evolution from
O-star through the Luminous Blue Variable (LBV)
phase to the Wolf-Rayet (WR) phase. O-stars have a
relatively small mass-loss rate ($\Mdot \simeq 10^{-6}$  to $10^{-5}$ \Msunyr) 
but high wind velocity ($\vinf \simeq$ few $10^{3}$ \kms), LBVs have a
high mass-loss rate (few $10^{-5}$ \Msunyr) but a small wind 
velocity (few $10^2$ \kms) and  WR-stars have a high
mass-loss rate (few $10^{-5}$ \Msunyr) 
and a high wind velocity (few $10^3$ \kms). The luminosity remains
approximately unchanged during these phases. 
The question is: {\it what causes these changes in mass-loss rate and in
wind velocity?}

The mass-loss rates of OB-stars and LBVs during their quiescent phase 
have been explained in terms of radiation driven 
winds, where the driving is done by
multitudes of spectral lines (Castor et al. \cite{cak75}, Pauldrach et al.
\cite{ppk86}, Vink et al. \cite{vink00}). 
Line driven winds are optically thin in their continuum.
Nugis \& Lamers (\cite{nl02}, hereafter NL) have shown that
the mass-loss rates of Wolf-Rayet stars can 
in principle
be explained by
optically thick radiation driven winds.
In this paper we explain the
large changes in \Mdot\ and \vinf\ 
in terms of transitions between four types of radiation driven
wind models, due to changes mainly
in the stellar parameters and to a smaller extent also  
in the surface abundance during the stellar evolution.
We discuss the properties of two types of
optically thin and two types of optically thick radiation driven wind
models, which we then
apply to nine characteristic massive stars with increasing evolution
stages.

\section{Evolutionary changes of the stellar 
parameters of massive stars}
\label{sec:evol}

The evolution of stars initially more massive than about 50
\Msun\ proceeds as follows (e.g. Maeder \& Meynet \cite{maed87}).
After the main sequence phase the
star expands and becomes a blue supergiant with a
radius of about $10^2$ \Rsun\ and with an enhanced He and N surface 
abundance. For
a reason that is poorly understood the star becomes an unstable
LBV with multiple outbursts 
(Humphreys \& Davidson \cite{hd94}; Leitherer \cite{leit97}).
Even during quiesence the mass-loss rates of LBVs are 
significantly higher than during the main
sequence phase but the wind velocity is much lower. After the LBV-phase
the star contracts and becomes a WR star with a N-rich (WN-type) and
later a  C,O-rich 
(WC-type) surface. The hydrostatic radius of the star is only a few 
\Rsun, but the high mass-loss rate produces an optically thick wind
with  a sonic point radii of about 15 to 30 \Rsun\ for hydrogen-rich
WNL-phase and about 1.5 to 10 \Rsun\ for the H-poor WNE/WC phase (see NL).
Very massive stars ($M>60~\Msun$) may skip the LBV phase (Bohannan \& Crowther 1999).

We adopt nine typical massive stars of about the same luminosity and initial
mass to characterize the changes in the stellar and wind parameters during the
evolution of a massive star. They represent different evolutionary phases.
The stars are listed in Table \ref{table1} in order of increasing
evolutionary stage: one Of-star ($\zeta$ Pup), one LBV (P Cyg), three
stars with spectra in between those of Of and late-WN (HD151804, HD
152408, HD152386), three WN-stars (WR~105, WR~136 and WR~139) and one
WC-star (WR~111). 
Notice the large changes in \Mdot\ and \vinf\ and the general increase of the 
atmospheric He/H-ratio and of the momentum transfer efficiency, 
$\eta=\Mdot \vinf /(L/c)$, with evolution stage.
The radii of the four genuine
WN and WC-stars in this table are the radii of the sonic point, derived
by NL, and the value of \Teff\ is at that radius.

\begin{table*}
\caption[]{Nine characteristic massive stars}
\begin{tabular}{cccccccccccl}
\hline\hline\noalign{\smallskip}
Name & Type & log $L$ & $R^{(1)}$ & \Teff & $M^{(2)}$ & \vesc
$^{(3)}$\ & $N_{\rm  He}/N_{\rm H}$ & \Mdot\ & \vinf\ & $\eta$ & Ref$^{(4)}$ \\
     &      & \Lsun   & \Rsun &  K  & \Msun & \kms   &       &
 \Msunyr & \kms &  &   \\ 
\hline\noalign{\smallskip}
$\zeta$ Pup & O4~If & 6.00 &  19 & 42~000 & 70 &953  & 0.15 &
3~$10^{-6}$ & 2250 & 0.32 &  PU  \\
 P Cyg & B1.5~Ia$^+$ & 5.86 & 76 & 19~300 & 23 & 223 & 0.30 &
2~$10^{-5}$ & 210  & 0.28    & PP  \\
HD151804& O8~Iaf     & 5.84 & 37 & 26~700 & 46 & 581 & 0.25 &
$1.2~10^{-5}$ & 1445 &1.24    & CB  \\
HD152408&  WN9ha   & 5.80 & 32 & 27~600 & 44 & 648 & 0.67 & 
$2.4~10^{-5}$ & 995  &1.79    & CB  \\
HD152386 & WN9ha     & 5.82 & 33 & 27~000 & 46 & 624 & 0.27 &
$2.7~10^{-5}$ & 1650 &3.45    & BC  \\      
 WR 105 & WN~9       &5.81  &26  & 32~100 & 22 & 412 & 0.44 &
2.8~$10^{-5}$ & 1200 &  2.6   & NL   \\ 
 WR 136 & WN~6b      & 5.73 & 4.6 & 73~000& 19 & 900 & 1.9 &
6.3~$10^{-5}$ & 1600 & 9.2 &   NL\\
WR 139 & WN~5      & 5.21  & 2.0 & 82~000 & 9.3 & 1129 & 5.0 &
0.9~$10^{-5}$ & 1800 & 4.9 &   NL \\ 
WR 111 & WC~5      & 5.31  & 2.3 & 81~000 & 10.6 & 1160 & $\infty$ &
1.0~$10^{-5}$ & 2415 & 5.8 & NL  \\
\hline
\end{tabular}

\noindent
$^{(1)}$ For WR-stars the mean sonic radius of the models A1 and B1 of
NL with the corresponding $\Teff =(L/(4 \pi \sigma R_{\rm s}^2))^{0.25}$ is
listed.\\
$^{(2)}$ Masses of O and WN9ha-stars were 
derived from evolutionary models.\\
$^{(3)}$ $\vesc$ is the effective escape velocity at the  radius $R$,
corrected for radiation pressure by electron scattering.\\
$^{(4)}$ PU = Puls et al. (\cite{puls96}); PP = Pauldrach \& Puls
(\cite{pp90});  CB = Crowther \& Bohannan (\cite{crow97});
BC = Bohannan \& Crowther (\cite{boh99});
NL = Nugis \& Lamers (\cite{nl02}). 
\label{table1}
\end{table*}

\section{Radiation driven wind models}
\label{sec3} 

\subsection{Optically thin line driven winds}
\label{sec3.1}

For line driven wind models, which are
optically thin in the continuum, the predicted terminal velocity \vinf\ is
\begin{equation}
\vinf \simeq C_{\rm fd} \sqrt{\alpha /(1-\alpha)}~ \vesc
\label{vinfCAK}
\end{equation}
(Castor et al. \cite{cak75}; Kudritzki et al. \cite{kud89}), 
where $\alpha$ is a force multiplier parameter with $\alpha \simeq 0.5$ to
0.7 for hot massive stars of $\Teff \simgreat 8000$ K and
\vesc\ is the effective escape velocity, 
i.e. corrected for radiation pressure by electron scattering.
Lamers et al. (\cite{lam95}) have shown that 
$ C_{\rm fd} \sqrt{\alpha /(1-\alpha)} \simeq 2.7$ if $\Teff \simgreat 
21~000$ K and 1.3 if $10~000 \simless \Teff \simless 21~000$ K.
Observations and theory both show that the mass-loss rate of a line
driven wind increases by about a factor 5 and the terminal
velocity decreases by about a factor two when the effective temperature of 
the star drops below about 21~000 K 
(Lamers et al. \cite{lam95}; Vink et al. \cite{vink99}). 
This is the {\it bi-stability jump},
 which is due to the change in ionization in the lower wind
layers near the sonic point (Vink et al. \cite{vink99}). 
Detailed calculations of line driven wind models with multiple scattering
by Vink et al. (\cite{vink01}) have shown that the mass-loss rate of 
galactic OB-stars in \Msunyr\ is

\begin{eqnarray}
\label{Vinkhot}
& \log \Mdot =   -6.86 + 2.194 \log (L/10^5) -1.313 \log (M/30) \nonumber\\ 
           & +  0.933 \log (\Teff/40\,000) 
            -  10.92 \{ \log(\Teff/40\,000)\}^2 
%
\end{eqnarray}
on the ``hot side'' of the 
jump, $\Teff \geq 21\,000$ K,  and

\begin{eqnarray}
\label{Vinkcool}
\log \Mdot  &=&   -6.39 + 2.210 \log (L/10^5) -1.339 \log (M/30) \nonumber\\ 
            &+&1.07 \log (\Teff/20\,000)
\end{eqnarray}
on the ``cool side'' of the jump,
$10\,000 \leq \Teff \leq 21\,000$ K, 
with $M$ and $L$ in solar units.
We will use Eqs. \ref{vinfCAK}, \ref{Vinkhot} and \ref{Vinkcool} 
to predict \vinf\ and \mdot\ for line driven winds.


\subsection{Optically thick radiation driven winds}
\label{sec3.2}

NL have shown that
in optically thick radiation driven winds 
the opacity has to increase outwards at the sonic point. 
They showed that 

\begin{equation}
\Mdot \approx C \frac{T_{\rm s}^4 R_{\rm s}^3 v_{\rm s}}{M}~ = ~4.66 \times 10^{-29}~ 
\frac{T_{\rm s}^{4.5} R_{\rm s}^{3}} {M}~ \sqrt{\frac{(1+\gamma)}{\mu}}
\label{Mdotthick}  
\end{equation}
in \Msunyr,
where $R_{\rm s}$ (in \Rsun) and $T_{\rm s}$ (in K) are the radius 
and temperature at the sonic
point, $M$ is in \Msun, 
$v_{\rm s}$
is the isothermal sound speed,
$\gamma$ is the mean number of free electrons per atom and $\mu$ is the 
mean atomic weight in atomic mass units.
We see that \Mdot\ of an optically thick radiation driven wind
is proportional to $T_{\rm s}^{4.5}$. At first sight this 
might suggest that an arbitrary high mass-loss rate 
can be reached by moving the sonic point deeper into the star
where the optical depth and the temperature are higher. However, this
is not the case, because the transition from subsonic to supersonic
velocity at the sonic point sets requirements for the opacity and 
its gradient. 
The opacity and its gradient at the sonic point are (see NL)

\begin{equation}
\chi_{\rm s} \simeq \frac{4 \pi c GM}{L}~,
\label{chis}
\end{equation}

\begin{equation}
\label{chigrad}
\left(\frac{d \chi}{dr}\right)_{\rm s} \simeq 
\chi_{\rm s} \frac {3 v_{\rm s}^{2}} {GM} = \chi_{\rm s} 
\frac{3 a_{1} T_{\rm s}}{GM}= \frac{12 \pi c a_1 T_{\rm s}}{L} >0~,
\end{equation}
where $a_1=k(\gamma+1)/(\mu m_{\rm u})$. 
Eqs. \ref{chis} and \ref{chigrad} imply that the transonic transition 
can only occur in the
layers where the opacity {\it increases} outwards and where it reaches
a value set by the luminosity and mass of the star (Eq. \ref{chis}). 
>From the 
OPAL-opacity tables (Iglesias \& Rogers \cite{ir96}) 
we find that this occurs only in limited temperature
regimes where $\chi(T)$ shows a bump. These regimes are in the ranges
of $156\,000 \leq  T_{\rm s} \leq 162\,000$ K, 
and $37\,000 \leq T_{\rm s} \leq
 71\,000$ K, where respectively a large and a small Fe-opacity peak 
occur (see NL).
We will use Eq. \ref{Mdotthick} 
to predict \Mdot\ for optically thick winds.
For $v_{\infty}$ we adopt the scaling predicted by the models.
We derived from WR models of NL that 

\begin{equation}
\vinf \simeq (2\pm0.5) v_{\rm esc}^{\rm s}
\label{vinf2}
\end{equation}
with $v_{\rm esc}^{\rm s}$ at the sonic point.
We will use this scaling law
to estimate \vinf\ for optically thick radiation driven
winds.
Notice that \vinf\ increases with decreasing 
sonic radius. 


\subsection{Four types of wind models}
\label{sec3.3}

The description above has shown that radiation driven winds from hot
stars come in four types:\\
(1) {\it line driven winds} which are optically thin in the continuum
for stars with $\Teff \geq 21~000$ K, i.e. on the hot side of the
bi-stability jump: ``line hot'' models.
For these winds we adopt Eqs. \ref{vinfCAK} and \ref{Vinkhot}. \\
(2) {\it line driven winds} 
for stars with $\Teff \leq 21~000$ K, i.e. on the cool side of the
bi-stability jump: ``line cool'' models.
We adopt the Eqs. \ref{vinfCAK} and \ref{Vinkcool} for these models.\\
(3) {\it ``thick cool'' continuum driven winds}, with the sonic point in the
temperature range of $38~000 < \Ts < 71~000$ K, where the small
opacity bump occurs (see NL).
For these stars we adopt Eqs. \ref{Mdotthick} and 
\ref{vinf2} with $\Ts=40~000$ and 70~000 K.\\
(4) {\it ``thick hot'' continuum driven winds}. 
For these stars we 
adopt Eqs. \ref{Mdotthick} and \ref{vinf2} with
$\Ts \simeq 160~000$ K.


\section{Predicted radiation driven mass-loss rates and velocities}
\label{Sec4}

We apply the predictions of the radiation driven wind models
to the nine stars. The resulting values of \Mdot\ and \vinf\ are
listed in Table 2 for four models: ``line cool'' or ``line hot'',
``thick cool''
with $\Ts = 40~000$ and $70~000$ K, 
and ``thick hot'' with $\Ts = 160~000$ K. 
For the line driven wind model of P Cyg we
adopt the predictions for the cool side 
of the bi-stability jump (``line cool'')
 whereas for the other stars we adopt the ``line hot'' models.

\begin{table*}
\caption[]{Predictions for line driven (optically thin continuum) 
and optically thick  winds versus observations. }
\begin{tabular}{lc|ccccc|cccl}
\hline\hline\noalign{\smallskip}
Star & Type &  & & log \Mdot &(\Msunyr)   &  & \vinf
& (\kms) &  & Best \\
 &     & thin & thick & thick & thick & obs & thin & thick &  obs & model
     \\
     & & line & cool  & cool  & hot   &     & line &
                &   & \\
     & &          &  40 kK & 70 kK & 160 kK &    &     & 
      &  &  \\
\hline{\smallskip}
 $\zeta$ Pup & O4~If &   -5.12&  -5.54& -4.44 & -2.83&  -5.52&  2573&  1906&  
 2250 & line hot \\
 P Cyg       & B1~Ia$^+$ &   -4.35&  -3.28&  -2.19& -0.58&  -4.70&   290&   446&   
 210 & line cool  \\
 HD~151804   & O8~Iaf    &   -5.71&  -4.51&  -3.42& -1.81&  -4.92&  1569&   1162&  
 1445& line/thick?\\
 HD~152408   & WN9ha     &   -5.68&  -4.73&  -3.65&-2.02&  -4.62&  1750&   1296&  
 955 & thick cool\\
 HD~152386   & WN9ha     &   -5.78&  -4.67& -3.57 & -1.96&  -4.56&  1685&   1248&  
 1650& thick cool\\
 WR~105      & WN~9      &   -5.09&  -4.68& -3.59 & -1.97&  -4.55&  1037&   824&   
 1200& thick cool\\
 WR~136      & WN~6b     &   -5.50&  -6.94& -5.84 & -4.22&  -4.20&  2430&  1800&  
 1600& thick hot\\
 WR~139      & WN~5      &   -6.50&  -7.73& -6.64 & -5.03&  -5.05&  3048&  2258&  
 1800 & thick hot\\
 WR~111      & WC 5      &   -6.33&  -7.70& -6.60 & -4.99&  -5.00&  3132&  2320&  
 2415  & thick hot\\    
\noalign{\smallskip}\hline
\end{tabular}
\label{table2}
\end{table*}

Comparing the predicted values of \Mdot\ and \vinf\ of the four
models with the observed values, we can determine which model 
fits best. In this comparison we have given more weight to the
mass-loss rate than to the terminal velocity, because the predicted
\vinf\ of the line driven models depends on the mass of the 
star, which is not well known, and \vinf\ of the optically thick
winds is not well predicted by NL models.  
For the star $\zeta$ Pup the models
``line hot'' and ``thick cool'' with $\Ts=40$
 kK predict almost the same mass-loss rates.
We adopt the ``line hot'' model, because this star is 
considered to be the prototype of a line driven wind
(e.g. Pauldrach et al. \cite{paul94}).
In the case of the O8~Iaf star HD~151804 \Mdot\
is in between the predicted values of the ``line hot'' and 
``thick cool'' (40 kK) models. 
The observed
value of \vinf\ agrees better with the ``line hot'' model. The wind of
this star may be of intermediate type. For the other stars the choice
of the best fitting model is quite obvious (last column of Table \ref{table2}).
Notice that for the ``thick cool'' models the observed values of \Mdot\
agree better with those predicted for $\Ts \simeq 40$ kK than for 
$\Ts \simeq 70$ kK.

\section{Discussion and conclusions}

>From the comparison between the predicted and the observed values of
\Mdot\ and \vinf, we can explain the changes in 
mass-loss rate and wind velocities during the evolution of massive
stars in terms of the four wind models.

\begin{enumerate}

\item  O-stars on the main sequence and shortly thereafter
have winds driven by lines on the hot side of the bi-stability jump.

\item  The winds of LBVs are radiation driven by lines.
The increase in \Mdot\ and the decrease in \vinf\ 
from O-star to LBV is due  to the crossing of the bi-stability jump.
(However, not all LBVs become cool enough to reach the bi-stability
jump: Leitherer \cite{leit97}; Lamers \cite{lam97})

\item  When the star has lost sufficient mass and the atmosphere 
has been He-enriched sufficiently to contract to the
WNL-phase, the wind becomes optically thick and the sonic point moves
into the region where the continuum opacity shows a small bump so
that it can initiate an optically thick wind. This results in a
(small) increase of \Mdot\ and a large increase in \vinf.
The characteristics of the star HD~151804 suggest that the transition
from line driven winds to ``thick cool'' is gradual.

\item  When the star looses more mass and the surface becomes H-poor
its wind may either stay ``thick  cool'' (at $\Ts \simeq 40$ kK) 
or become ``thick hot''. 
In the first case the star may appear  as a H-poor WN7 or WN8-star
(not studied here). In the 
latter case the star appears as a WNE-star, similar to WR~136 and
WR~139. In both cases the star has a high \Mdot\ and a high \vinf.
The transition from a ``thick cool'' to ``thick hot'' wind
is determined by the variation of $\chi_{\rm s}$ (see Eq. \ref{chis}). 
During the WNE-phase the $M/L$-ratio increases
(Schaerer \& Maeder \cite{schaer92}). 
When $\chi_{\rm s}$ increases to values above $\simeq$0.5 cm$^2$g$^{-1}$,
the sonic point {\it has} to move to high temperature regime, because
such a high value of $\chi_{\rm s}$ 
is reached only near the main iron opacity peak (NL).
When the star evolves directly from Of to WNLh both the $M/L$-ratio
and $\chi_{\rm s}$ decrease.

\item 
It is difficult to predict the dependence of $\dot{M}$
on $L$ for optically thick wind models accurately because 
$R_{\rm s}$ and hence also $\Teff(R_{\rm s})$ is not well known.
The wind models of NL for WNE/WCE stars predict that
$R_{\rm s} \propto L^{0.7}$ and because $M \propto L^{0.6}$ 
(Schaerer \& Maeder \cite{schaer92}),  
it follows that  $\dot{M} \propto \Rs^3/M \propto L^{1.5}$ (Eq.  
\ref{Mdotthick}). This dependence agrees 
well with the empirical relation derived by Nugis \& Lamers (\cite{nl00})
for WN stars. For the WC stars Nugis \& Lamers (2000) found
empirically that
$\dot{M} \propto L^{0.84}$ but with strong dependence on chemical
composition. On the other hand, in a recent study of LMC WC-stars 
Crowther et al. (A\&A in press) found a strong
dependence of $\dot{M}$ on $L$ ($\dot{M} \propto L^{1.38}$)
which agrees well with our predicted dependence for 
optically ``thick hot'' wind models.

\item
We find that the changes in \Mdot\ and \vinf\ during the
  evolution of massive stars are mainly due to changes in the stellar
  parameters and to a lesser degree to changes in the surface composition.

\end{enumerate}

We have shown that the changes in $\Mdot$ and $\vinf$ during the
evolution of the massive stars from O-star to WN-star are due to the
adjustment of the wind to the changing conditions,
 mainly the $M$, $L$, $R$ and surface composition. We have not
 explained ``how'' these changes occur.
The transition from line driven 
winds to ``thick cool'' winds is most likely due to the formation of a
bump in the opacity curve when the He/H ratio increases. 
This transition 
can occur gradually because the temperature of the sonic point of a 
``line hot'' wind overlaps with the range for ``thick cool'' winds.
The transition from ``thick cool'' to ``thick hot'' is probably due to
the fact that the hydrostatic radius of a WR-star shrinks and the
$M/L$-ratio increases when the
luminosity of the star decreases (Schaerer \& Maeder \cite{schaer92}). 
This transition cannot be gradual,
because the sonic point temperature in these models is very
different, so the wind must be restructured during this transition.

We point out that at present, 
the optically thick wind models do not provide accurate predictions for
$\vinf$, so we adopted an empirical
scaling law. However, Schmutz (\cite{schmutz97}) has shown that there is
sufficient driving in the supersonic part of the winds of WR-stars 
to explain the observed high values of \vinf.

%
\acknowledgements
We thank an unkown referee for positive and constructive comments.
This work was supported by the Estonian Science Foundation grant No. 5003. 
T.N. is grateful to the Netherlands School of Astronomy (NOVA) for a
travel grant.
%

%

%

\end{document}